\documentclass{article} 
\usepackage{iclr2027_conference,times}

\usepackage{amsmath,amsfonts,bm}

\def\eqref#1{equation~\ref{#1}}

\def\1{\bm{1}}

\DeclareMathAlphabet{\mathsfit}{\encodingdefault}{\sfdefault}{m}{sl}
\SetMathAlphabet{\mathsfit}{bold}{\encodingdefault}{\sfdefault}{bx}{n}

\usepackage{hyperref}
\usepackage{url}
\usepackage{graphicx}

\usepackage{bm}
\renewcommand{\vec}[1]{\bm{#1}}

\title{On the stability of learning}

\author{%
  John J. Vastola \\
Department of Neurobiology, Harvard Medical School \\
Department of Psychology and Center for Brain Science, Harvard University \\
Kempner Institute for the Study of Natural and Artificial Intelligence, Harvard University \\
  \texttt{john\_vastola@hms.harvard.edu} 
}

\iclrfinalcopy 
\begin{document}

\maketitle
\pagestyle{plain}

\begin{abstract}
Learning algorithms are usually analyzed under the assumption that informative gradients are continuously available. But this is not always true---learning signals in biological neural networks, for example, may be sparse, local, and only intermittently available at any given synapse. In this work, we show that if gradient information is noisy and only intermittently available, learning may be unstable. In particular, even if average learning dynamics converge to the correct optimum, variance in learning may be asymptotically infinite. We explore this point in a number of analytically tractable settings, including single-layer and multi-layer linear networks, and derive explicit conditions under which learning is stable. These conditions complement, but are conceptually distinct from, existing conditions related to the stability of stochastic gradient descent. Our work demonstrates that unbiased credit assignment is not enough for reliable learning; sufficiently frequent and informative feedback is also required, which yields an additional constraint on models of biologically plausible learning.
\end{abstract}

\section{Introduction}

The claim that the brain does not perform backpropagation \citep{rumelhart_learning_1986} rests in part on the idea that gradient information may be difficult to communicate. In particular, the gradient of an objective function with respect to any given synaptic weight depends on the synaptic weights of a large number of downstream neurons, and it is not obvious in general how such non-local information could be reliably communicated \citep{crick_recent_1989,lillicrap_backpropagation_2020}. This issue has motivated a substantial body of work seeking alternative and more biologically plausible learning rules \citep{lillicrap_random_2016,scellier_equilibrium_2017}, as well as schemes for approximately implementing backpropagation in specific contexts \citep{sacramento_dendritic_2018,whittington_theories_2019,millidge_predictive_2022}.
It remains unknown if and how this information is communicated, but it is generally agreed that communicating it accurately would be difficult given locality constraints.

In this paper, we consider a related but conceptually distinct issue. If gradient information \textit{can} be communicated, it is not at all obvious that it can be communicated \textit{reliably}. It is plausible that a scheme which delivers the correct updates \textit{on average} only delivers them occasionally, to different neurons at different times, and perhaps somewhat noise-corrupted. How do learning dynamics change when gradients are noisy and only intermittently available? We argue that learning changes substantially, and in fact can become unstable if updates are not delivered sufficiently frequently.

This idea is interesting for a few reasons. First, it implies that credit assignment which is correct \textit{on average} is insufficient, since there remains the additional obstacle that weight updates cannot be too variable. Second, instability is not a generic consequence of noisy weight dynamics, but rather depends crucially on the multiplicative, sparse nature of the noise introduced by assuming intermittent gradient arrivals.

We show that the severity of this issue can be reduced by a few obvious noise mitigation strategies, but that the presence of the issue (i.e., learning becoming unstable if feedback is sufficiently sparse) survives. This problem exists in both linear and nonlinear networks; in shallow and deep networks; and after averaging parameters or recent gradients. The details of the issue change, but the core finding remains: an informative signal must arrive sufficiently often for learning to remain stable.

We proceed as follows. We begin by motivating the simplest version of our model. Next, by analyzing it in detail, we show that there exist gradient-variance-dependent stability issues that are exacerbated as informative feedback becomes less frequent. We then discuss simple examples of this issue near the fixed points of more complex networks, and the extent to which basic noise mitigation strategies address it. 

\section{Sparse gradient arrivals can amplify errors}
\label{sec:sparse}

\begin{figure}[ht!]
\begin{center}
\includegraphics[width=\textwidth]{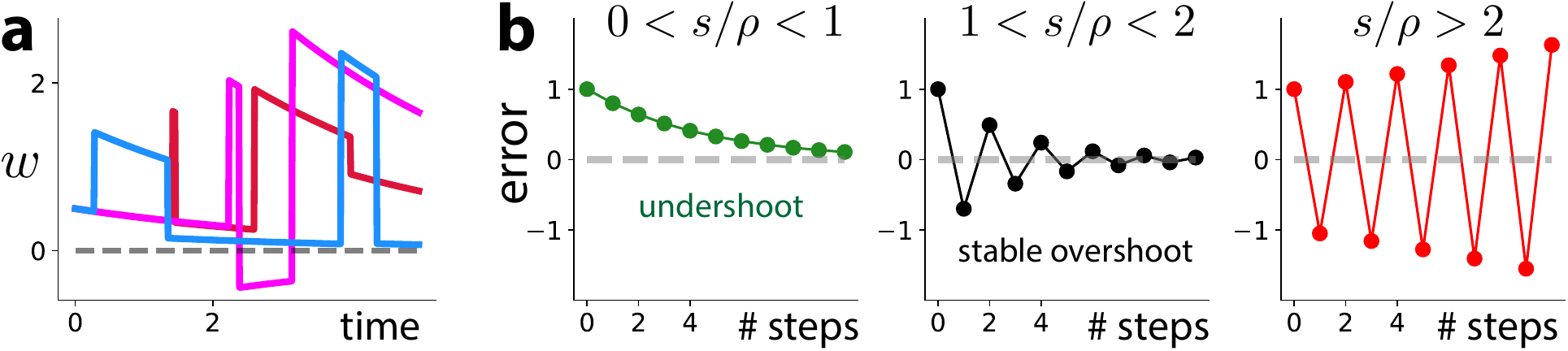}
\end{center}
\caption{\textbf{Sparse gradient arrivals: motivation and intuition.} (\textbf{a}) A simple example of sparse gradient arrivals: a weight receives intermittent updates, in between which the regularization term produces relaxation toward zero. Three example stochastic simulations are shown. (\textbf{b}) Error as a function of number of arrivals in a toy model. If $s/\rho < 1$, error always decreases (left); if $1 < s/\rho < 2$, error alternates in sign but decreases in magnitude (center); and if $s/\rho > 2$, learning becomes unstable, and error both alternates in sign and increases in magnitude (right).}
\label{fig:shooting}
\end{figure}

Suppose we are interested in updating a set of weights $\vec{w} \in \mathbb{R}^N$ to optimize an objective
\begin{equation} \label{eq:basic}
\tilde{J}(\vec{w}) := J(\vec{w}) + \frac{\gamma}{2} \Vert \vec{w} \Vert_2^2
\end{equation}
where $J$ is a task-related objective and the second term regularizes $\vec{w}$. We explicitly separate $J$ and the regularization term because their biological interpretation in what follows is different. In the gradient-flow limit typical of many theoretical treatments of learning \citep{saxe_linear_2013,bordelon_2020,domine_lazy_2025}, gradient descent involves updates of the form
\begin{equation} \label{eq:gd_naive}
\frac{d}{dt} \vec{w} = \eta \left[ - \frac{\partial J}{\partial \vec{w}} - \gamma \vec{w} \right]
\end{equation}
where $\eta > 0$ is the learning rate and $t \in [0 ,\infty)$ is the amount of elapsed training time. We will investigate the consequences of assuming $w_i$ only has \textit{intermittent} access to the correct gradient. We do this as follows. We model sparse gradient arrivals via a Poisson process \citep{kingman1992poisson}: we assume that each synapse has a fixed probability per unit time of receiving a noise-corrupted version of the relevant gradient information. In doing things this way, we ignore potential temporal structure in gradient arrivals and correlations between the information arriving at different synapses; accounting for these factors will likely complexify the resulting model, and more importantly, make the associated stability issues worse rather than better.

\paragraph{Poisson gradient arrivals assumption.} To model temporally sparse updates, we assume gradients are coupled to a \textit{Poisson process} with mean arrival rate $\eta \rho$, where $\rho > 0$ is the non-dimensionalized \textit{arrival rate}. As $\rho$ increases, neurons tend to receive more gradient arrivals per learning time scale $1/\eta$. If we denote a realization of this Poisson process by $\vec{r}(t) \in \mathbb{R}^N$, our model corresponds to the replacement
\begin{equation} \label{eq:replace}
- \frac{\partial J}{\partial \vec{w}} \to  \vec{r}(t) \odot \frac{1}{\rho} \left[ - \frac{\partial J}{\partial \vec{w}}  + \vec{\xi} \right] 
\end{equation}
where $\odot$ denotes the Hadamard (or \textit{elementwise}) product. The quantity $\vec{\xi} \sim \mathcal{N}( \vec{0}, \sigma^2_{\text{obs}} \vec{I})$ represents \textit{observation noise}. Our main motivation for including it is to illustrate how, unlike sparse gradient arrivals, this kind of noise is comparatively benign. 

According to this prescription, weight changes due to gradient arrivals have mean and covariance
\begin{equation}
\mathbb{E}[ \Delta \vec{w} ] = - \eta \Delta t \frac{\partial J}{\partial \vec{w}} \hspace{1in} \text{Cov}( \Delta w_i, \Delta w_j ) = \frac{\eta \Delta t}{\rho} \  \delta_{i j} \left[ \left( \frac{\partial J}{\partial \vec{w}}  \right)^2 + \sigma^2_{\text{obs}}    \right] \ .
\end{equation}

\paragraph{Useful intuition.} We can make a simple heuristic argument to explain why sparse gradient arrivals can destabilize learning. Consider gradient descent (Eq. \ref{eq:gd_naive}) with only one weight and no regularization ($\gamma = 0$) on a mean-squared error objective
\begin{equation}
J := \frac{s}{2} (w - w^*)^2 \ .
\end{equation}
For Poisson gradient arrivals, weights do not change until there is an arrival, so learning effectively occurs in discrete time. The error $e_n := w_n - w^*$ after $n$ gradient arrivals evolves according to the recurrence relation
\begin{equation}
e_{n + 1} = \left( 1 - \frac{s}{\rho} \right) e_n \ .
\end{equation}
These dynamics yield a geometric series with $e_n = ( 1 - s/\rho)^n e_0$ which has three regimes. For a sufficiently large arrival rate ($s/\rho < 1$), error consistently decreases and asymptotically approaches zero (Fig. \ref{fig:shooting}b, left). For a moderately large arrival rate ($1 < s/\rho < 2$), weight dynamics overshoots its target and the error alternates between positive and negative; it still asymptotically approaches zero (Fig. \ref{fig:shooting}b, center). If the arrival rate is too small ($s/\rho > 2$), learning becomes unstable, and error both alternates in sign and increases in magnitude (Fig. \ref{fig:shooting}b, right). This indicates that, at least for this toy model, the ratio $s/\rho$ between input variance and the gradient arrival rate controls whether learning is stable. Hence, the arrival rate $\rho$ must be larger than
\begin{equation} \label{eq:rho_c_naive}
\boxed{ \ \rho > \rho_c := \frac{s}{2} \ } \ .
\end{equation}
We will see that more complex models exhibit surprisingly similar behavior, and in particular lead to similar expressions for the critical arrival rate $\rho_c$.

\section{Instability in a linear model}
\label{sec:linear}

\begin{figure}[ht!]
\begin{center}
\includegraphics[width=\textwidth]{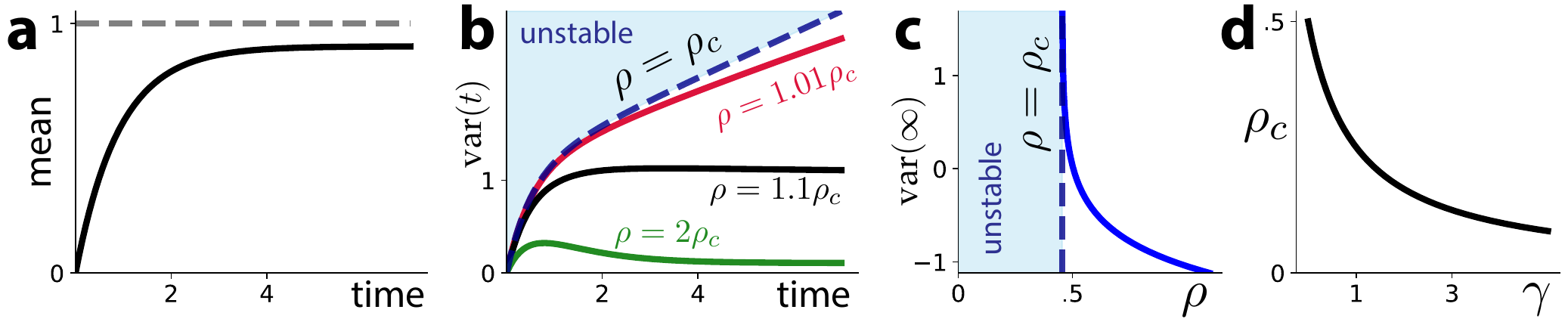}
\end{center}
\caption{\textbf{Instability in a linear model.}  (\textbf{a}) Average dynamics are unconditionally stable and do not depend on the arrival rate $\rho$. (\textbf{b}) Output variance converges more slowly as $\rho$ decreases and diverges to infinity when $\rho < \rho_c$. (\textbf{c}) The steady state output variance diverges like $1/(\rho - \rho_c)$. (\textbf{d}) $\rho_c$ decreases as $\gamma$ increases, so one way to improve stability is to increase the regularization strength.}
\label{fig:linear}
\end{figure}

In this section, we study instability due to sparse gradient arrivals in a tractable linear model, in part because we think it reveals the essence of the phenomenon. Consider an objective
\begin{equation}
\tilde{J} =  \mathbb{E}_{\vec{x}}\left\{ \frac{\Vert \vec{w}^T \vec{x} - (\vec{w}^*)^T \vec{x} \Vert_2^2}{2} \right\} + \frac{\gamma}{2} \Vert \vec{w} \Vert_2^2 = \frac{1}{2} (\vec{w} - \vec{w}^*)^T \vec{S} (\vec{w} - \vec{w}^*)  + \frac{\gamma}{2} \Vert \vec{w} \Vert_2^2
\end{equation}
where the input $\vec{x}$ is assumed to have mean zero and covariance $\vec{S} \in \mathbb{R}^{N \times N}$. The negative gradient of $J$ equals $\vec{S} (\vec{w}^* - \vec{w})$, so our prescription yields dynamics
\begin{equation} \label{eq:model_linear}
\frac{d}{dt} \vec{w} = \vec{r} \odot \frac{1}{ \rho} [ \vec{S} (\vec{w}^* - \vec{w}) + \vec{\xi} ] - \eta \gamma \vec{w} \ .
\end{equation}
The average $\bar{\vec{w}} := \mathbb{E}[ \vec{w} ]$ evolves according to standard-looking linear dynamics
\begin{equation}
\frac{d}{dt} \bar{\vec{w}} = \eta \left[ \vec{S} (\vec{w}^* - \bar{\vec{w}})  - \gamma \bar{\vec{w}} \right]
\end{equation}
which approaches the fixed point at $\vec{w}' := (\vec{S} + \gamma \vec{I})^{-1} \vec{S} \vec{w}^*$ exponentially quickly. On the other hand, the dynamics of the covariance matrix $\vec{C} := \mathbb{E}[ (\vec{w} - \bar{\vec{w}}) (\vec{w} - \bar{\vec{w}})^T]$ are substantially more interesting (see Appendix \ref{app:linear}):
\begin{equation} \label{eq:cov_dyn_linear}
\boxed{ \ \frac{1}{\eta} \frac{d}{dt} \vec{C} = - 2 \gamma \vec{C} -  \vec{C} \vec{S} -  \vec{S} \vec{C} + \frac{1}{\rho}  \sigma^2_{\text{obs}} \vec{I} + \frac{1}{\rho} \ \text{diag}( \bar{g}_i^2 + [ \vec{S} \vec{C} \vec{S} ]_{i i}  ) \ } \ .
\end{equation}
We can assess stability by determining the conditions under which $\lim_{t \to \infty} \text{tr}[\vec{C}(t)]$ is finite. While this is difficult to do analytically in general, it is possible to do if we make at least one additional assumption about $\vec{S}$: either that (i) $\vec{S}$ is \textit{diagonal}, or that (ii) $\vec{S}$ is \textit{circulant}.

\subsection{Stability given a diagonal input covariance}

If $\vec{S}$ is diagonal, Eq. \ref{eq:cov_dyn_linear} simplifies since $[ \vec{S} \vec{C} \vec{S} ]_{ii} = S_{i i}^2 C_{i i}$. Let us simplify notation by writing $c_i := C_{ii}$ and $s_i := S_{ii}$. The diagonal elements of $\vec{C}$ then evolve according to
\begin{equation} 
\frac{1}{\eta} \frac{d}{dt} c_i = \frac{1}{\rho} ( \sigma^2_{\text{obs}} + \bar{g}_i^2 )   - \left( 2 \gamma c_i +  2 s_i - \frac{s_i^2}{\rho} \right) c_i  \ .
\end{equation}
These dynamics are linear, and whether $c_i$ converges to a stable value or diverges to infinity depends on the coefficient of $c_i$---that is, it must be negative. For learning to be stable, the coefficients of all $c_i$ must be negative; hence, it must be true that
\begin{equation} \label{eq:rho_c_diag}
\boxed{ \ \rho > \rho_c^i := \frac{s_i^2}{2 (s_i + \gamma)} \ }
\end{equation}
for all $i$. Equivalently, $\rho > \rho_c := \text{max}_i \rho_c^i$, which depends on the largest input variance $s_i$. Note that this condition reduces to the critical arrival rate associated with our naive model (Eq. \ref{eq:rho_c_naive}) when we take $\gamma \to 0$. 

As $\rho$ decreases, $c_i$ converges increasingly slowly, and below $\rho_c$ it becomes unstable (Fig. \ref{fig:linear}b). The steady state diagonal covariance elements diverge like $1/(\rho - \rho_c^i)$ (Fig. \ref{fig:linear}c), or more quantitatively,
\begin{equation}
c_{i} = \frac{1}{\rho - \rho_c^i} \frac{\sigma^2_{\text{obs}} + \bar{g}_i^2}{2 (s_i + \gamma)} \ .
\end{equation}
Our formula indicates that the critical arrival rate threshold $\rho_c$ can be reduced in two ways: one can either decrease input variance or increase the regularization strength $\gamma$ (Fig. \ref{fig:linear}d). The latter method recalls the standard bias-variance tradeoff \citep{bishop2006pattern}: we can decrease the (infinite) variance at the cost of increasing bias, since increasing $\gamma$ makes $\vec{w}$ more different from the target weight $\vec{w}^*$.

\subsection{Stability given a circulant input covariance}

Suppose instead that $\vec{S}$ is \textit{circulant}. Then $\vec{S}$ can be diagonalized via a discrete Fourier basis, i.e., $\vec{S} = \vec{F} \vec{\Lambda} \vec{F}^{\dag}$ (see Appendix \ref{app:linear}). The arrival-related term can be written in terms of the covariance matrix in that basis, $\tilde{\vec{C}} := \vec{F}^{\dag} \vec{C} \vec{F}$, since
\begin{equation}
\frac{1}{N} \sum_i [ \vec{S} \vec{C} \vec{S} ]_{ii} = \frac{1}{N} \sum_q s_q^2 \tilde{C}_{q q} \ .
\end{equation}
In this case, different $\tilde{C}_{q q}$ do not decouple, but it is still possible to straightforwardly characterize dynamics. After some algebra, we obtain a critical arrival rate 
\begin{equation} \label{eq:rho_c_circ}
\boxed{ \ \rho > \rho_c := \frac{1}{2 N} \sum_q \frac{s_q^2}{s_q + \gamma}  \ } \ .
\end{equation}
Unlike the previous result, this one involves the entire spectrum of $\vec{S}$ (although it reduces to the previous one when all $s_q$ are the same). Still, the result conveys the same qualitative message: $\rho_c$ increases as the input variance increases, and decreases as the regularization strength $\gamma$ increases.

\section{Instability in a nonlinear network}

\begin{figure}[ht!]
\begin{center}
\includegraphics[width=\textwidth]{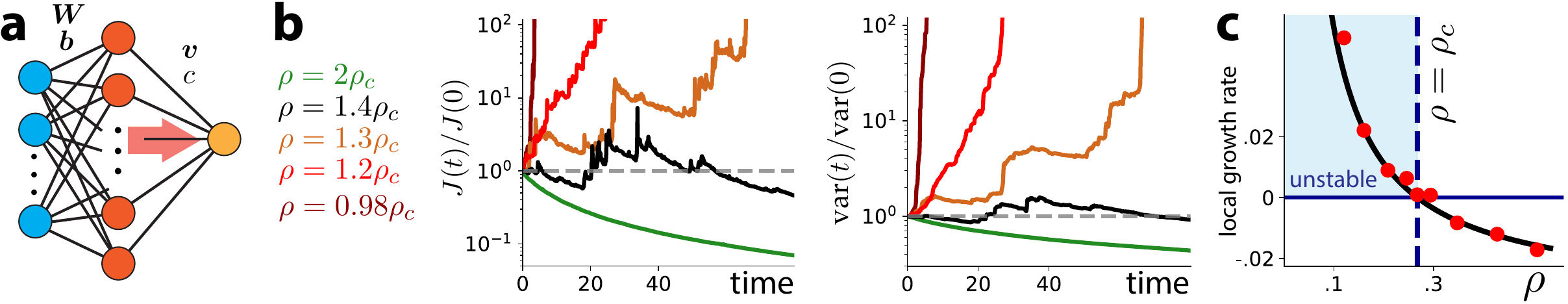}
\end{center}
\caption{\textbf{Instability in a two-layer network.} (\textbf{a}) We trained a two-layer network to reproduce the output of an identical teacher network. To ensure our linear model provides a (locally) reasonable description of its dynamics, we initialized it near the teacher's weights. (\textbf{b}) Normalized objective and output variance throughout learning. Note that for $\rho < \rho_c$ (dark red), $J$ blows up. For $\rho$ somewhat above the threshold (red, orange), learning still looks unstable on this finite time scale. Learning only appears stable well beyond the threshold (green). (\textbf{c}) Instantaneous growth rate of the covariance dynamics near the fixed point. The black line indicates the linear model's prediction and the red dots indicate numerical measurements. Below $\rho = \rho_c$, this local growth rate can be positive, which indicates instability.}
\label{fig:nonlinear}
\end{figure}

One virtue of our linear model analysis is that it implies sparse gradient arrivals pose problems for much more general networks. Near a stable fixed point $\vec{w}^*$ of any well-behaved objective $J$, 
\begin{equation}
J(\vec{w}) \approx J(\vec{w}^*) + \frac{1}{2} (\vec{w} - \vec{w}^*)^T \vec{H} (\vec{w} - \vec{w}^*)
\end{equation}
where $\vec{H}$ denotes the (positive definite) Hessian of $J$ at the fixed point $\vec{w}^*$. Hence, gradient descent dynamics are approximately linear near this fixed point, and if we make the identification $\vec{S} \to \vec{H}$, they are well-described by our linear model (Eq. \ref{eq:model_linear}).

\paragraph{Example: a two-layer network.} Although the above mathematical argument is fairly persuasive, it remains helpful to see an explicit nonlinear example where sparse gradient arrivals destabilize learning. For concreteness' sake, we trained a simple two-layer network to reproduce the output of a teacher network with the same architecture (Fig. \ref{fig:nonlinear}a). We initialized the student network near the teacher's weights and assessed how changing the arrival rate $\rho$ affected learning. For simplicity and to speed up computation, only the hidden layer's biases $\vec{b}$ and the output parameters $\vec{v}$ and $c$ were trained; we found that this did not affect our conclusions, however.

We numerically computed the Hessian around the teacher network's parameters in order to estimate the critical arrival rate $\rho_c$ below which learning ought to be unstable. For $\rho < \rho_c$, we found that the loss function and output variance blew up as learning proceeded (Fig. \ref{fig:nonlinear}b, c; the dark red line indicates $\rho = 0.98 \rho_c$). In fact, we found that $\rho$ marginally above the critical arrival rate ($\rho = 1.2 \rho_c$, red; $\rho = 1.3 \rho_c$, orange) also exhibited unstable learning dynamics on the finite time scales we examined. Learning only became stable, with a smoothly-decreasing loss, when $\rho$ was much larger than $\rho_c$ ($\rho = 2 \rho_c$, green). 

We also tested whether the instantaneous change in the output variance (or the ``local growth rate''), which is predicted by Eq. \ref{eq:cov_dyn_linear}, empirically describes small-time changes in output variance. We find that this is indeed the case, and that the local growth rate is positive below $\rho = \rho_c$ (Fig. \ref{fig:nonlinear}c).

\section{Instability in a deep linear network}
\label{sec:deep}

\begin{figure}[ht!]
\begin{center}
\includegraphics[width=0.8\textwidth]{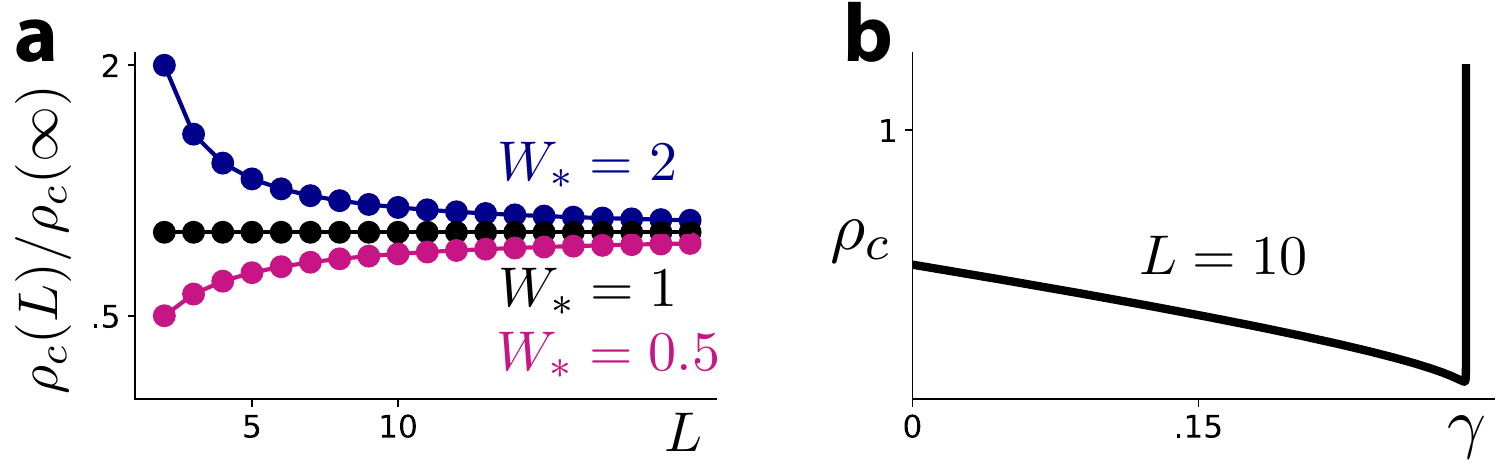}
\end{center}
\caption{\textbf{Depth benignly modulates stability in a deep linear network.} (\textbf{a}) $\rho_c$ can increase, decrease, or remain constant with $L$ depending on the value of the target $W_*$. In all cases, the difference between $\rho_c(L)$ and its asymptotic value $\rho_c(\infty)$ is not substantial. (\textbf{b}) $\rho_c$ is non-monotonic as a function of $\gamma$ ($L = 10$ shown), but this is mainly because sufficiently high regularization destabilizes the nontrivial fixed point.}
\label{fig:deep}
\end{figure}

Are deeper networks less stable in the sense we study here? To partially address this question in an analytically tractable setting, we consider an extremely simple deep linear network with $L \geq 1$ layers whose hidden `layers' are just scalars:
\begin{equation}
y = w_L \cdots w_1 x
\end{equation}
where $w_i$ is the scalar weight associated with the $i$-th `layer', $x$ is a scalar input, and $y$ is a scalar output. Although deep linear networks are admittedly much simpler than their nonlinear counterparts, they have the virtue that they are analytically tractable, and they recapitulate a surprising amount of the phenomenology associated with more realistic networks \citep{saxe_linear_2013,saxe_semantic_2019,domine_lazy_2025}.

Consider a teacher-student setting in which our goal is to match $y$ to a target output $y_* := W_* x$:
\begin{equation}
J = \mathbb{E}_x\left\{   \frac{( y - y_*)^2}{2} \right\} = \frac{v}{2} (  w_1 \cdots w_L - W_* )^2
\end{equation}
where we have assumed that $x$ has mean zero and variance $v > 0$. Although the gradient descent learning dynamics of this network are generally nonlinear and complicated, we can analyze it around the \textit{symmetric fixed point} with $w_1 = \cdots = w_L = q$. We will study the impact of depth on the stability of dynamics near this fixed point. Note that we are studying the fixed point of the regularized objective $\tilde{J}$ (see Eq. \ref{eq:basic}), not $J$.

Define the quantity $h := v q^{2 L - 2}$. The Hessian $\vec{H}$ of $J$ about the symmetric fixed point has two eigenvalues (see Appendix \ref{app:deep}):
\begin{equation}
h_{\perp}=  \gamma \hspace{0.2in} (\text{with multiplicity } L-1) \hspace{0.5in} h_{\parallel} = \gamma + L (h - \gamma) \hspace{0.2in} (\text{with multiplicity } 1) \ .
\end{equation}

Since $\vec{H}$ is circulant, we can immediately reuse the result from Sec. \ref{sec:linear} to find that
\begin{equation}
\boxed{ \ \rho_c(L) = \frac{1}{2 L} \left[ \frac{h_{\parallel}^2}{h_{\parallel} + \gamma} + (L - 1) \frac{h_{\perp}^2}{h_{\perp} + \gamma} \right] = \frac{1}{2 L} \left[ \frac{[ \gamma + L (h - \gamma) ]^2}{ L (h - \gamma) + 2 \gamma} + \frac{(L - 1)}{2} \gamma  \right] \ } \ .
\end{equation}
In the case of no regularization ($\gamma = 0$), we have the simpler critical arrival rate
\begin{equation}
\boxed{ \ \rho_c(L) = \frac{h}{2} = \frac{v}{2} W_*^{2 - 2/L}  \ } \ .
\end{equation}
Interestingly, $\rho_c$ only depends benignly on $L$, and moreover has a well-behaved infinite-depth ($L \to \infty$) limit (Fig. \ref{fig:deep}a). There is an apparent instability as $\gamma$ increases, since $\rho_c$ diverges to infinity at a finite value of $\gamma$ (Fig. \ref{fig:deep}b), but this is misleading; it occurs not because covariance dynamics become unstable, but because it is no longer optimal to match $W_*$ when there is too much regularization.

At least in this case, depth does not exacerbate stability issues. In some sense this should be unsurprising, since a result that suggests $\rho_c$ scales poorly with depth would seem inconsistent with biology, whose networks are often deep and highly recurrent; that is, learning is hard, but it should not be impossible.

\section{Pooling parameters can improve stability}
\label{sec:pool}

\begin{figure}[ht!]
\begin{center}
\includegraphics[width=\textwidth]{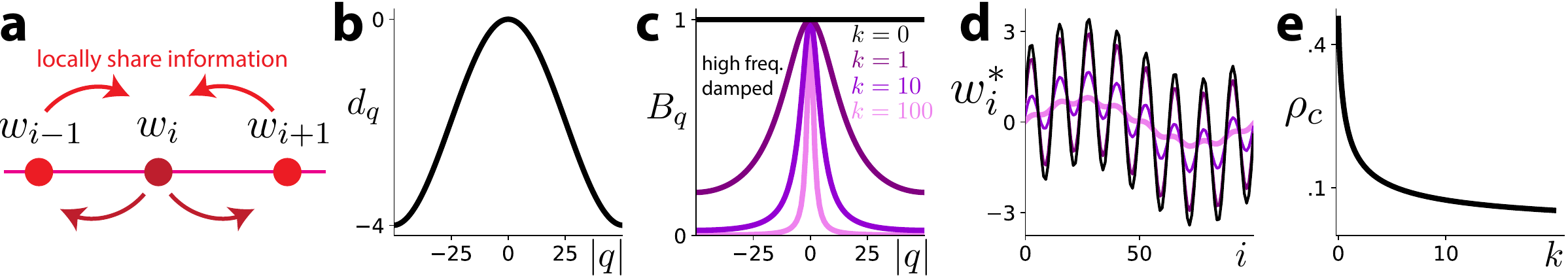}
\end{center}
\caption{\textbf{Pooling parameters can improve stability.} (\textbf{a}) If nearby synapses store similar information, neurons may benefit from sharing information with their spatial neighbors. Pictured is a line of synaptic weights; arrows indicate bidirectional information sharing. (\textbf{b}) The eigenvalues $d_q$ of $\vec{D}$ are indexed by spatial frequencies $|q|$, and $d_q$ decreases as $|q|$ increases. (\textbf{c}) The $q$-dependence of $d_q$ causes higher spatial frequencies of the fixed point to be damped, and the effect size increases as the amount of diffusion $k$ increases (more purple $\to$ higher $k$). (\textbf{d}) One effect of this frequency-dependence is to smooth the fixed point of the average dynamics. In the pictured example, $w_i^*$ is highly oscillatory (black), but dynamics converge to a smoothed version of it (shades of purple). (\textbf{e}) $\rho_c$ decreases as $k$ increases, so although parameter-pooling produces bias, it also improves stability.}
\label{fig:pool}
\end{figure}

One approach to improving stability is to average over parameters, since if nearby neurons have similar weights, it may make sense for them to share information (Fig. \ref{fig:pool}a). Spatially proximal neurons are known to communicate in a variety of non-synaptic ways, including through gap junctions \citep{shimizu_gap_2013} and extracellular vesicles \citep{nieves_torres_inter-neuronal_2023}, and although the computational role of these features is unknown, it is plausible that they share computationally relevant information, including regarding synaptic weights. Parameter-pooling inevitably increases bias, but this tradeoff may be worth it if learning is otherwise unstable.

Assume that weights are arranged along a one-dimensional ring, so that $w_i$ is `near' $w_{i+1}$ and $w_{i-1}$. Also assume that $\vec{S}$ is circulant. Consider a slight modification of Eq. \ref{eq:model_linear} which incorporates parameter sharing via a `diffusion' matrix $\vec{D}$ which satisfies $[ \vec{D} \vec{w} ]_i = w_{i-1} - 2 w_i + w_{i+1}$:
\begin{equation}
\frac{d}{dt} \vec{w} = \vec{r} \odot \frac{1}{ \rho} [ \vec{S} (\vec{w}^* - \vec{w}) + \vec{\xi} ] - \eta \gamma \vec{w} + \eta k \vec{D} \vec{w} \ .
\end{equation}
Here, $k \geq 0$ quantifies the strength of diffusion. This model reduces to the one studied in Sec. \ref{sec:linear} when $k = 0$, but otherwise behaves differently. Its average dynamics read
\begin{equation}
\frac{1}{\eta} \frac{d}{dt} \vec{w} =  \vec{S} (\vec{w}^* - \vec{w})  -  \gamma \vec{w} + k \vec{D} \vec{w}
\end{equation}
and, in the Fourier basis that diagonalizes $\vec{D}$ and $\vec{S}$, the components of the fixed point are
\begin{equation}
w_q^{\text{FP}} := \frac{s_q}{s_q + \gamma - k d_q} w^*_q = B_q w^*_q \ .
\end{equation}
The prefactor $B_q$, which controls the asymptotic bias in learning, is highly spatial-frequency-dependent: because $d_q$ is more negative for large frequencies $|q|$ (Fig. \ref{fig:pool}b), $B_q$ is closer to zero for large $|q|$, and the effect size increases as the diffusion strength $k$ increases (Fig. \ref{fig:pool}c). This implies that what is learned is a spatially smoothed version of the target (Fig. \ref{fig:pool}d).  

Meanwhile, following algebra similar to what we encountered in Sec. \ref{sec:linear} (see Appendix \ref{app:pool}), covariance dynamics are stable provided that a stability condition
\begin{equation}
\boxed{ \ \rho > \rho_c := \frac{1}{2  N} \sum_q \frac{s_q^2}{s_q + \gamma - k d_q}  \ }
\end{equation}
holds. Since $\rho_c$ decreases as $k$ increases, pooling increases stability  (Fig. \ref{fig:pool}e).

\section{Averaging over gradient arrivals can improve stability}
\label{sec:avggrad}

\begin{figure}[ht!]
\begin{center}
\includegraphics[width=\textwidth]{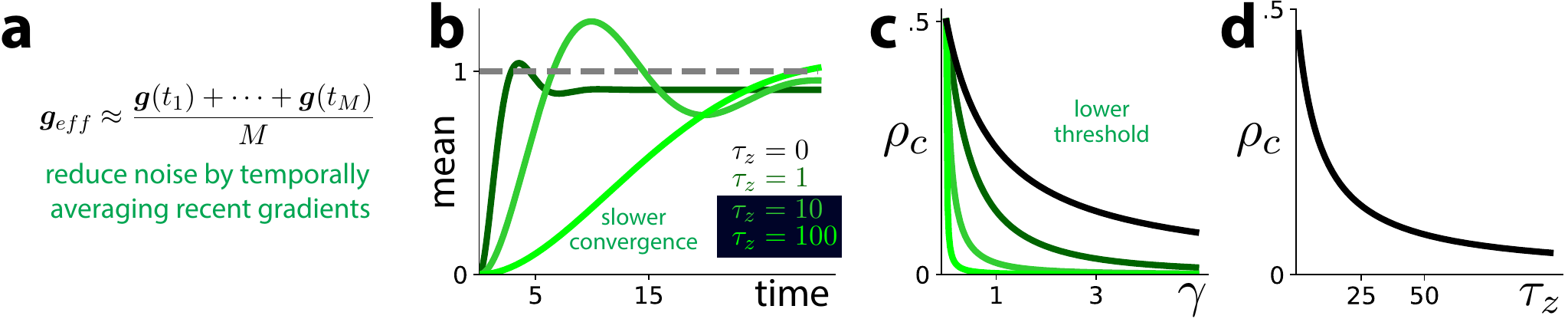}
\end{center}
\caption{\textbf{Averaging over gradient arrivals can improve stability.} (\textbf{a}) Intuitively, one can reduce variance by performing a temporal average over recent gradient arrivals. (\textbf{b}) Mean dynamics are linear but not monotonic. As the gradient-averaging time scale $\tau_z$ increases, they converge more slowly. (black: $\tau_z = 0$; lightest green: $\tau_z = 100$) (\textbf{c}, \textbf{d}) Increasing $\tau_z$ improves stability since $\rho_c$ decreases as long as $\gamma > 0$. But if $\gamma = 0$, there is no stability benefit. }
\label{fig:avggrad}
\end{figure}

Another plausible way to improve stability is to effectively reduce the variance of gradient arrivals by not learning \textit{directly} from the available gradient signal, but from a temporal average of recent arrivals (Fig. \ref{fig:avggrad}a). We can concretely instantiate this idea by assuming that each parameter $w_i$ has a matching accumulator variable $z_i$ which leaky-integrates recent gradients on a time scale $\tau_z > 0$:
\begin{equation}
\begin{split}
\frac{d}{dt} \vec{z} &= \vec{r} \odot \frac{1}{\eta \tau_z \rho} [ \vec{S} (\vec{w}^* - \vec{w}) + \vec{\xi} ]  - \frac{1}{\tau_z} \vec{z} \\
\frac{d}{dt} \vec{w} &= \eta ( \vec{z} - \gamma \vec{w} ) \ .
\end{split}
\end{equation}
By construction, the fixed point of average $\vec{w}$ dynamics is the same as before, although its transient behavior is slightly different (Fig. \ref{fig:avggrad}b). Significantly, though, the addition of the accumulator variable $\vec{z}$ greatly changes covariance dynamics. The covariances $\vec{Z} := \text{Cov}(\vec{z})$, $\vec{X} := \text{Cov}(\vec{z}, \vec{w})$, and $\vec{W} := \text{Cov}(\vec{w})$ evolve according to the ODEs (see Appendix \ref{app:avggrad})
\begin{equation}
\begin{split}
\frac{d}{dt} \vec{Z} &= - \frac{2}{\tau_z} \vec{Z} - \frac{1}{\tau_z} ( \vec{X} \vec{S} + \vec{S} \vec{X}^T )   + \frac{1}{\eta \rho \tau_z^2} \text{diag}[ g_i(\bar{\vec{w}})^2 + [\vec{S} \vec{W} \vec{S}]_{ii} + \sigma^2_{\text{obs}}  ] \\
\frac{d}{dt} \vec{X} &= \eta \vec{Z}- \left( \eta \gamma + \frac{1}{\tau_z} \right) \vec{X} - \frac{1}{\tau_z} \vec{S} \vec{W} \\
\frac{d}{dt} \vec{W} &=  \eta \left[ \vec{X} + \vec{X}^T - 2 \gamma \vec{W}  \right] \ .
\end{split}
\end{equation}
As before, it is much easier to analyze these dynamics if we make specific assumptions about $\vec{S}$. We will again assume that $\vec{S}$ is either diagonal or circulant.

\subsection{Stability given a diagonal input covariance}

If $\vec{S}$ is diagonal, the dynamics of $( Z_{ii}, X_{ii}, W_{ii})$ for any fixed $i$ are independent of those for $i' \neq i$, and are characterized by a $3 \times 3$ matrix
\begin{equation}
\vec{M} := \begin{pmatrix}
- \frac{2}{\tau_z} & - \frac{2 s_i}{\tau_z}  & \frac{s_i^2}{\eta \rho \tau_z^2}\\
\eta &  -  \left( \eta \gamma + \frac{1}{\tau_z} \right)  & - \frac{s_i}{\tau_z} \\
0 & 2 \eta  &  - 2 \eta \gamma
\end{pmatrix} \ .
\end{equation}
Covariance dynamics are stable provided that $\vec{M}$ has only negative eigenvalues, and this holds unless $\rho$ is sufficiently small. We can find where the critical value $\rho_c^i$ is by computing where the determinant of $\vec{M}$ becomes zero. This yields a stability condition
\begin{equation}
\boxed{ \ \rho > \rho_c^i := \frac{s_i^2 }{2( 1 + \eta \gamma \tau_z) (  s_i + \gamma )} \ }
\end{equation}
analogous to what we discovered before. It reduces to our earlier result (Eq. \ref{eq:rho_c_diag}) in the $\tau_z \to 0$ limit, but otherwise produces a lower critical value of $\rho$ (Fig. \ref{fig:avggrad}c, d). Relatedly, we obtain steady state weight variances
\begin{equation}
 W_{i i} = \frac{1}{\rho - \rho_c^i} \frac{\sigma^2_{\text{obs}} + \bar{g}_i^2}{2 (s_i + \gamma) ( 1 + \gamma \eta \tau_z)}
\end{equation}
which again diverge like $1/(\rho - \rho_c^i)$. It is important to emphasize that this gain in stability does not come for free: increases in $\tau_z$ can also greatly slow down the convergence of average learning dynamics (Fig. \ref{fig:avggrad}b, light green line).

\subsection{Stability given a circulant input covariance}

For circulant $\vec{S}$, after some algebra, we obtain a result
\begin{equation}
\boxed{ \ \rho > \frac{1}{2 N (1 + \eta \gamma \tau_z)} \sum_q \frac{s_q^2}{s_q + \gamma}    \ }
\end{equation}
which is again similar to what we found before (c.f. Eq. \ref{eq:rho_c_circ}), and reduces to it in the $\tau_z \to 0$ limit.

\section{Discussion}

Learning is difficult, and in this work we have identified and analyzed yet another reason why. It is not enough to perform accurate credit assignment \textit{on average} if the relevant learning signal is highly variable, since this high variance can produce unstable learning. The source of this variance, sparse gradient arrivals, is distinct from other types of noise relevant to learning (e.g., noise due to using stochastic rather than full-batch gradient descent), and its effects are apparently more insidious. 

Because any well-behaved network is approximately linear near a stable fixed point, one expects this issue to be generic rather than architecture-dependent. We considered three ways of addressing the problem---\textit{regularization}, \textit{parameter-sharing}, and \textit{gradient-averaging}---and each reduces the severity of the issue but involves tradeoffs. Regularization and parameter-sharing reduce the (infinite) variance at the expense of bias; moreover, parameter-sharing may only be natural when there is specific (e.g., topographic) structure according to which spatial relationships correlate with relationships between synaptic weights. Gradient-averaging does not produce bias, but does slow down learning. It is plausible that all of these mechanisms, and perhaps others, are utilized by biological networks to avoid the kind of instability we study here.

\paragraph{Connection to analyses of stochastic gradient descent.} Standard analyses of stochastic gradient descent \citep{wu_how_2018,chaudhuri_sgd_2022} involve a mathematical idea similar to the one we study here, whose content is especially similar to the naive model we considered in Sec. \ref{sec:sparse}: if the learning rate or step size is taken to be too large, errors will accumulate and learning will become unstable. In other ways, the issue is qualitatively different. Our model is defined in continuous- rather than discrete-time, and moreover involves a nontrivial interaction between discrete and stochastic gradient arrivals on the one hand, and continuous relaxation (due to regularization) and accumulation (in the model considered in Sec. \ref{sec:avggrad}) on the other. Moreover, the biological interpretation of sparse arrivals---which can be viewed as related to error-communication difficulties rather than simply an overall learning rate---is conceptually distinct. Finally, average dynamics are unconditionally stable.

\paragraph{Speculation regarding biological means of variance reduction.} If biological systems do indeed confront the problem of potential instability due to sparse gradient (or more generally, learning signal) arrivals, it is plausible that evolution has developed mechanisms to partially combat it. There is biological precedent for each of the three mechanisms---regularization, parameter-pooling, and gradient-averaging---we considered. It is known that the dendritic spines, which are sites of synaptic plasticity, turn over on relatively short time scales, e.g., on the order of one or two weeks in the hippocampus \citep{attardo_impermanence_2015}. This continual `forgetting' of synaptic weights can be modeled in the same way we have incorporated regularization. Our gradient-averaging model can be viewed as maintaining a kind of \textit{eligibility trace} of learning-related information, and there is mounting evidence that such traces exist and play a major role in biological learning \citep{he_distinct_2015,gerstner_eligibility_2018}. Parameter-pooling has not been directly observed to our knowledge, but as we have mentioned, there are plausible means by which it could occur.

\paragraph{Limitations and future work.} For clarity and tractability, we have mainly analyzed simple linear models here. More interesting behavior may be possible in less linear regimes, e.g., away from a stable fixed point. Another major assumption we made is that sparse gradients arrive independently at each synapse according to a Poisson process. In reality, the delivery of gradient information to any given neuron is likely correlated with the delivery of gradients to other neurons (e.g., because the information is coming from roughly the same source). Although details associated with this issue likely differ for more general models, our observations suggest that it is worth further study. Putative methods of credit assignment must produce both good average performance and stable variance.

\clearpage

\subsection*{AI use statement}

In this work, we used generative AI tools for brainstorming, literature search, to assist with checking the mathematical correctness of our results, and to assist with code generation. The author has checked all text and results and takes full responsibility for them.

\bibliography{stab_bib}
\bibliographystyle{iclr2027_conference}

\clearpage

\appendix
\section{Analysis of linear model}
\label{app:linear}

In this appendix, we analyze the behavior of the linear model discussed in Sec. \ref{sec:linear} of the main text. Recall that it has
\begin{equation} \label{eq:model_linear_app}
\frac{d}{dt} \vec{w} = \vec{r} \odot \frac{1}{ \rho} [ \vec{S} (\vec{w}^* - \vec{w}) + \vec{\xi} ] - \eta \gamma \vec{w} 
\end{equation}
where $\vec{r}$ is distributed according to a Poisson arrival process with rate $\eta \rho$ and $\xi \sim \mathcal{N}(0, \sigma^2_{\text{obs}})$ is observation noise associated with a gradient arrival. The master equation that defines how the distribution $P(\vec{w} | t)$ of weight values evolves in time is
\begin{equation} \label{eq:master}
\frac{\partial P}{\partial t} =  \eta \frac{\partial}{\partial \vec{w}} \cdot \left[  \gamma \vec{w} P  \right] + \eta \rho \sum_i  \int_{-\infty}^{\infty} d\xi  \ \mathcal{N}(\xi; 0, \sigma^2_{\text{obs}}) \ 
\left[ P(\vec{w} - (1/\rho)[g_i(\vec w)+\xi]\vec{e}_i | t) - P(\vec{w} | t) \right]
\end{equation}
where $\vec{e}_i$ denotes the standard basis vector with a $1$ as its $i$-th component and zeros elsewhere. $g_i(\vec{w})$ is shorthand notation for the $i$-th component of $\vec{S} (\vec{w}^* - \vec{w})$, the true gradient.

\subsection{Derivation of mean dynamics}

Let us first derive the dynamics of the mean weight $\bar{\vec{w}} := \mathbb{E}[ \vec{w} ]$. The master equation (Eq. \ref{eq:master}) has the form of a Fokker-Planck equation plus the Poisson arrival term. The regularization term contributes the expected $- \eta \gamma \bar{\vec{w}}$. We can determine the contribution of the Poisson arrival term by computing its average over $\vec{w}$. Note that
\begin{equation}
\begin{split}
& \ \eta \rho \int \vec{w} \left\{  \int_{-\infty}^{\infty} d\xi  \ \mathcal{N}(\xi; 0, \sigma^2_{\text{obs}}) \ 
\left[ P(\vec{w} - (1/\rho) [g_i(\vec w)+\xi]\vec{e}_i | t) - P(\vec{w} | t) \right] \right\} \ d\vec{w} \\
=& \ \eta \rho \int  \ \mathcal{N}(\xi; 0, \sigma^2_{\text{obs}}) \ 
\left[ \vec{w} P(\vec{w} - (1/\rho)[g_i(\vec w)+\xi]\vec{e}_i | t) - \vec{w} P(\vec{w} | t) \right]  \ d\vec{w} \\
=& \ \eta \rho \int  \ \mathcal{N}(\xi; 0, \sigma^2_{\text{obs}}) \ 
\left[  \vec{w} + (1/\rho)(g_i(\vec w)+\xi) \vec{e}_i  - \vec{w}  \right] P(\vec{w} | t)   \ d\vec{w} \\
=& \ \eta \ \mathbb{E}[ g_i(\vec{w}) \vec{e}_i ] \ .
\end{split}
\end{equation}
Summing over $i$, we find that average dynamics are
\begin{equation}
\boxed{ \ \frac{d}{dt} \bar{\vec{w}} = \eta \left[ \vec{S} (\vec{w}^* - \bar{\vec{w}})  - \gamma \bar{\vec{w}} \right]  \ } \ ,
\end{equation}
i.e., they are unbiased.

\subsection{Derivation of covariance dynamics}

Let us now derive the dynamics of the covariance matrix $\vec{C} := \text{Cov}(\vec{w})$. Again, the regularization term contributes to its dynamics in the usual way; the hard part is to determine how the Poisson arrival term contributes. 

The average of the Poisson arrival term times $\vec{w} \vec{w}^T$ equals
\begin{equation}
\begin{split}
& \ \eta \rho \int \vec{w} \vec{w}^T \left\{  \int_{-\infty}^{\infty} d\xi  \ \mathcal{N}(\xi; 0, \sigma^2_{\text{obs}}) \ 
\left[ P(\vec{w} - (1/\rho) [g_i(\vec w)+\xi]\vec{e}_i | t) - P(\vec{w} | t) \right] \right\} d\vec{w} \\
=& \ \eta \rho \int  \ \mathcal{N}(\xi; 0, \sigma^2_{\text{obs}}) \ 
\left[ \vec{w} \vec{w}^T P(\vec{w} - (1/\rho) [g_i(\vec w)+\xi]\vec{e}_i | t) - \vec{w} \vec{w}^T P(\vec{w} | t) \right] \ d\vec{w} \\
=& \ \eta \rho \int  \ \mathcal{N}(\xi; 0, \sigma^2_{\text{obs}}) \ 
\left\{ \ [ \vec{w} + (1/\rho) (g_i(\vec w)+\xi) \vec{e}_i ] [ \vec{w} +(1/\rho)  (g_i(\vec w)+\xi) \vec{e}_i ]^T - \vec{w}  \vec{w}^T \  \right\} P(\vec{w} | t) \\
=& \ \eta \ \mathbb{E}[ ( \vec{w} \vec{e}_i^T + \vec{e}_i \vec{w}^T) g_i(\vec{w}) + (1/\rho) ( g_i(\vec{w})^2 + \sigma^2_{\text{obs}}  )\vec{e}_i \vec{e}_i^T ] \ .
\end{split}
\end{equation}
Summing over $i$, this becomes
\begin{equation}
\begin{split}
& \ \eta \sum_i \mathbb{E}[ ( \vec{w} \vec{e}_i^T + \vec{e}_i \vec{w}^T) g_i(\vec{w}) + (1/\rho) ( g_i(\vec{w})^2 + \sigma^2_{\text{obs}}  )\vec{e}_i \vec{e}_i^T ] \\
=& \ \eta \ \mathbb{E}[ \vec{w} (\vec{w}^* - \vec{w})^T \vec{S} + \vec{S} (\vec{w}^* - \vec{w}) \vec{w}^T  + (1/\rho) \text{diag}( g_i(\vec{w})^2 + \sigma^2_{\text{obs}}  ) ] \ .
\end{split}
\end{equation}
Meanwhile, the time derivative of the mean product $\bar{\vec{w}} \bar{\vec{w}}^T$ equals
\begin{equation}
\begin{split}
\frac{d}{dt} ( \bar{\vec{w}} \bar{\vec{w}}^T ) &= \left( \frac{d}{dt} \bar{\vec{w}}  \right) \bar{\vec{w}}^T  + \bar{\vec{w}}  \left( \frac{d}{dt} \bar{\vec{w}} \right)^T \\
&= \eta \left[  \vec{S} (\vec{w}^* - \bar{\vec{w}}) - \gamma \bar{\vec{w}} \right] \bar{\vec{w}}^T + \eta \bar{\vec{w}} \left[  \vec{S} (\vec{w}^* - \bar{\vec{w}}) - \gamma \bar{\vec{w}} \right]^T \ .
\end{split}
\end{equation}
Putting these results together, we find that $\vec{C} = \mathbb{E}[ \vec{w} \vec{w}^T] - \bar{\vec{w}} \bar{\vec{w}}^T$ evolves according to
\begin{equation}
\frac{1}{\eta} \frac{d}{dt} \vec{C} = - 2 \gamma \vec{C} - \vec{C} \vec{S} - \vec{S} \vec{C} + (1/\rho) \ \text{diag}( \mathbb{E}[g_i(\vec{w})^2] + \sigma^2_{\text{obs}}  ) \ .
\end{equation}
This result can be slightly simplified using the fact that
\begin{equation}
\mathbb{E}[ g_i(\vec{w})^2 ] = \bar{g}_i^2 + [ \vec{S} \vec{C} \vec{S} ]_{i i} \ ,
\end{equation}
i.e., using the fact that the squared gradient can be written as a sum of `bias' and `variance' terms. Here $\bar{g}_i$ is shorthand notation for $g_i(\bar{\vec{w}})$. We conclude that
\begin{equation} \label{eq:dyn_cov_app}
\boxed{ \ \frac{1}{\eta} \frac{d}{dt} \vec{C} = - 2 \gamma \vec{C} -  \vec{C} \vec{S} -  \vec{S} \vec{C} + \frac{1}{\rho}  \sigma^2_{\text{obs}} \vec{I} + \frac{1}{\rho} \ \text{diag}( \bar{g}_i^2 + [ \vec{S} \vec{C} \vec{S} ]_{i i}  ) \ } \ .
\end{equation}
Analysis of Eq. \ref{eq:dyn_cov_app} is cumbersome for general $\vec{S}$, but greatly simplifies for specific choices of $\vec{S}$. In the next two subsections, we perform a stability analysis of these covariance dynamics assuming $\vec{S}$ is either diagonal or circulant.

\subsection{Stability analysis for diagonal $S$}

Assume $\vec{S}$ is diagonal. Let $c_i := C_{ii}$ and $s_i := S_{i i}$, and collect both into the vectors $\vec{c}$ and $\vec{s}$. Eq. \ref{eq:dyn_cov_app} implies that the diagonal entries of $\vec{C}$ satisfy
\begin{equation} \label{eq:dyn_cov_app_diag}
\begin{split}
\frac{1}{\eta} \frac{d}{dt} \vec{c} &= - 2 \gamma \vec{c} - 2  \vec{S} \vec{c}  + \frac{1}{\rho}  \sigma^2_{\text{obs}} \vec{1} + \frac{1}{\rho} \ [ \vec{g} + \vec{S}^2 \vec{c} ] \\
 &=   \frac{1}{\rho} [  \sigma^2_{\text{obs}} \vec{1} +  \vec{g}] - [ 2 \gamma \vec{I} + 2  \vec{S} - \frac{1}{\rho} \vec{S}^2 ]  \vec{c} 
\end{split}
\end{equation}
where we use $\vec{g}$ to denote the vector whose components are $\bar{g}_i^2$ and $\vec{1}$ to denote the vector whose components are all ones. 

\paragraph{Stability.} Because this ODE is linear, it is clear that stability requires that
\begin{equation}
2 \gamma + 2  s_i - \frac{1}{\rho} s_i^2 > 0
\end{equation}
for all $i$. Rearranging, this implies that we require 
\begin{equation}
\boxed{ \ \rho > \rho_c^i := \frac{s_i^2}{2 (s_i + \gamma)} \ }
\end{equation}
for all $i$, or equivalently that 
\begin{equation}
\boxed{ \ \rho > \rho_c := \frac{s_{\text{max}}^2}{2 (s_{\text{max}} + \gamma)} \ }
\end{equation}
where $s_{\text{max}}$ is the largest input variance $s_i$.

\paragraph{Steady state covariance.} Setting the left-hand side of Eq. \ref{eq:dyn_cov_app_diag} to zero, we find that the steady state covariance is
\begin{equation}
\boxed{ \ C_{i i} = \frac{1}{\rho - \rho_c^i} \frac{\sigma^2_{\text{obs}} + g_i(\bar{\vec{w}}^*)^2}{2 (s_i + \gamma)} \ } \ ,
\end{equation}
where $\bar{\vec{w}}^{*}$ denotes the fixed point of mean dynamics, provided that $\rho > \rho_c$.

\subsection{Stability analysis for circulant $S$}

Assume $\vec{S}$ is circulant. Importantly, circulant matrices can be diagonalized via (discrete) Fourier basis functions. Concretely, let $\vec{F}$ denote the $N \times N$ matrix whose components are
\begin{equation}
F_{j q} = \frac{1}{\sqrt{N}} e^{i q j} \ .
\end{equation}
Then we can write $\vec{S} = \vec{F} \vec{\Lambda} \vec{F}^{\dag}$, where $\vec{\Lambda}$ is a diagonal matrix containing the eigenvalues of $\vec{S}$. Moreover, if we define the transformed covariance $\tilde{\vec{C}} := \vec{F}^{\dag} \vec{C} \vec{F}$, then 
\begin{equation}
\frac{1}{N} \sum_i [ \vec{S} \vec{C} \vec{S} ]_{ii} = \frac{1}{N} \sum_q s_q^2 \tilde{C}_{q q} \ ,
\end{equation}
where $s_q$ here denotes the $q$-th eigenvalue of $\vec{S}$. The virtue of this result, and hence the reason the circulant assumption is useful, is that the above quantity is independent of the index $i$.

Combining this result with Eq. \ref{eq:dyn_cov_app}, we find that the diagonal entries $\tilde{C}_{q q}$ of the transformed covariance satisfy
\begin{equation} \label{eq:dyn_cov_circ_raw}
\frac{1}{\eta} \frac{d}{dt} \tilde{C}_{qq} = - 2 \gamma \tilde{C}_{qq} - 2 s_q  \tilde{C}_{q q} + \frac{1}{\rho}  \sigma^2_{\text{obs}}  + \frac{1}{\rho} \ ( A_{q q} +\frac{1}{N} \sum_{q'} s_{q'}^2 \tilde{C}_{q' q'}  ) 
\end{equation}
where $\tilde{\vec{A}} := \vec{F}^{\dag} \text{diag}( \bar{g}_i^2 ) \vec{F}$.

\paragraph{Stability.} Define $c_q := \tilde{C}_{q q}$ and collect the $c_q$ into the vector $\vec{c}$. Also, collect the $\tilde{A}_{q q}$ into $\vec{a}$. Eq. \ref{eq:dyn_cov_circ_raw} can be written in the form
\begin{equation} \label{eq:dyn_cov_circ_raw_vec}
\begin{split}
\frac{1}{\eta} \frac{d}{dt} \vec{c} &= - 2 \gamma \vec{c} - 2 \vec{\Lambda} \vec{c} + \frac{1}{\rho}  \sigma^2_{\text{obs}} \vec{1}  + \frac{1}{\rho} \ ( \vec{a} +\frac{1}{N} \vec{1} \vec{1}^T \vec{\Lambda}^2 \vec{c})  \\
&= \frac{1}{\rho} ( \sigma^2_{\text{obs}} \vec{1} + \vec{a})  - \left[ 2 \gamma \vec{I} + 2 \vec{\Lambda}  - \frac{1}{\rho N}  \vec{1} \vec{1}^T \vec{\Lambda}^2  \right] \vec{c}  \\
&= \frac{1}{\rho} ( \sigma^2_{\text{obs}} \vec{1} + \vec{a})  - 2 \gamma \left[  \vec{I} + (1/\gamma) \vec{\Lambda}  - \frac{1}{2 \rho N \gamma}  \vec{1} \vec{1}^T \vec{\Lambda}^2  \right] \vec{c}  \ .
\end{split}
\end{equation}
From this result, it is clear that the stability of $\vec{c}$ dynamics depends on the matrix 
\begin{equation}
\vec{M} := \vec{I} + \frac{1}{\gamma} \vec{\Lambda} - \frac{1}{2 \rho N \gamma} \vec{1} \vec{1}^T \vec{\Lambda}^2 \ .
\end{equation}
This a diagonal matrix plus a rank-one perturbation. We can use the matrix determinant lemma to compute its determinant:
\begin{equation}
\begin{split}
\det(\vec{M}) &= \det(  \vec{I} + \frac{1}{\gamma} \vec{\Lambda}   ) \left[  1 - \frac{1}{2 \rho N \gamma} \vec{1}^T \vec{\Lambda}^2 ( \vec{I} + \frac{1}{\gamma} \vec{\Lambda} )^{-1}  \vec{1}  \right] \\
&= \det(  \vec{I} + \frac{1}{\gamma} \vec{\Lambda}   ) \left[  1 - \frac{1}{2 \rho N \gamma} \sum_q \frac{s_q^2}{s_q/\gamma + 1}  \right] \ .
\end{split}
\end{equation}
For sufficiently high $\rho$, it is clear that this determinant is positive, which indicates stability. We can identify the stability threshold by determining where the determinant is zero; since the eigenvalues of $\vec{S}$ are all positive, this can only happen when the bracketed factor equals zero. 

We conclude that stability requires
\begin{equation}
\boxed{ \ \rho > \rho_c := \frac{1}{2 N} \sum_q \frac{s_q^2}{s_q + \gamma}  \ }  
\end{equation}
since beyond this value of $\rho$, the determinant is negative. 

\paragraph{Steady state covariance.} The above covariance dynamics can be rewritten as 
\begin{equation} \label{eq:dyn_cov_circ_raw_vec2}
\begin{split}
\frac{1}{\eta} \frac{d}{dt} \vec{c} &= \frac{1}{\rho} ( \sigma^2_{\text{obs}} \vec{1} + \vec{a})  - 2 \gamma \vec{M} \vec{c}  \ ,
\end{split}
\end{equation}
so its steady state occurs at
\begin{equation}
\vec{c}^* := \frac{1}{2 \rho \gamma} \vec{M}^{-1} ( \sigma^2_{\text{obs}} \vec{1} + \vec{a})  \ .
\end{equation}
This can be simplified if we write out $\vec{M}^{-1}$ explicitly. By the Sherman-Morrison formula,
\begin{equation}
\begin{split}
\vec{M}^{-1} &= (  \vec{I} + \vec{\Lambda}/\gamma)^{-1} + \frac{1}{2 \rho N \gamma} \frac{(  \vec{I} + \vec{\Lambda}/\gamma)^{-1} \vec{1} \vec{1}^T \vec{\Lambda}^2 (  \vec{I} + \vec{\Lambda}/\gamma)^{-1} }{1 - \frac{1}{2 \rho N \gamma} \vec{1}^T \vec{\Lambda}^2 (  \vec{I} + \vec{\Lambda}/\gamma)^{-1} \vec{1}} \\ 
&= (  \vec{I} + \vec{\Lambda}/\gamma)^{-1} + \frac{1}{2  N \gamma} \frac{(  \vec{I} + \vec{\Lambda}/\gamma)^{-1} \vec{1} \vec{1}^T \vec{\Lambda}^2 (  \vec{I} + \vec{\Lambda}/\gamma)^{-1} }{\rho - \rho_c} \ .
\end{split}
\end{equation}
Although the expression we obtain is more complicated than in the case of diagonal $\vec{S}$, we still see a divergence that goes like $1/(\rho - \rho_c)$ as $\rho$ approaches the critical arrival rate from above.

\clearpage

\clearpage

\section{More details on the deep linear network}
\label{app:deep}

In this appendix, we provide a more thorough account of the deep linear network analysis described in Sec. \ref{sec:deep}. We consider a scalar deep network with $L \geq 2$ `layers':
\begin{equation}
y = w_L \cdots w_1 x \ .
\end{equation}
Here, $x$ is the input, $w_i$ is the weight associated with the $i$-th `layer', and $y$ is the output. $x$ is assumed to have mean zero and variance $v$.

We assume this network is trained via gradient descent to match a target output $y_* := W_* x$ which also depends linearly on its input. The corresponding objective equals
\begin{equation}
\tilde{J} = J + \sum_i \frac{\gamma}{2} w_i^2 =  \frac{v}{2} (  w_1 \cdots w_L - W_* )^2 + \sum_i \frac{\gamma}{2} w_i^2 \ .
\end{equation}

\paragraph{Fixed point assumption.} Although learning dynamics are generally nonlinear, they are approximately linear near a fixed point. For simplicity, we will consider the symmetric fixed point with 
\begin{equation}
w_1 = \cdots = w_L = q \ ,
\end{equation}
where $q$ is chosen to minimize the objective, i.e., 
\begin{equation}
q := \text{argmin}_q \frac{v}{2} (  q^L - W_* )^2 + \frac{L \gamma}{2} q^2 \ .
\end{equation}
Since $q$ minimizes $\tilde{J}$, $d\tilde{J}/dq = 0$, so the condition
\begin{equation}
\boxed{ \ v (W_* - q^L ) q^{L-2} = \gamma  \ }
\end{equation}
holds.

\paragraph{Hessian near fixed point.} To characterize linear dynamics near the symmetric fixed point, we need to compute the corresponding $L \times L$ Hessian matrix. Note that the first derivatives at an arbitrary point are
\begin{equation}
\frac{\partial J}{\partial w_i} = v ( w_1 \cdots w_L - W_*) \frac{W}{w_i}
\end{equation}
where we define $W := w_1 \cdots w_L$. The Hessian's components at an arbitrary point are
\begin{equation}
\begin{split}
H_{i i} &= v \frac{W^2}{w_i^2} \\
H_{i j} &= v\frac{W^2}{w_i w_j} + v ( W - W_*) \frac{W}{w_i w_j} \ .
\end{split}
\end{equation}
At the symmetric fixed point,
\begin{equation}
\begin{split}
H_{i i} &= v q^{2 L-2} \\
H_{i j} &= v q^{2 L-2} + v ( q^L - W_*) q^{L-2} \ .
\end{split}
\end{equation}
To ease notation, define $h := v q^{2 L - 2}$. This result implies that we can write the Hessian matrix at the symmetric fixed point as
\begin{equation}
\vec{H} = \gamma \vec{I} + (h - \gamma) \vec{1} \vec{1}^T \ ,
\end{equation}
i.e., it is a diagonal matrix plus a rank-one `correction'. We immediately learn that $\vec{H}$ has two eigenvalues: one associated with the correction, and another eigenvalue (with multiplicity $L-1$) associated with the orthogonal directions:
\begin{equation}
h_{\perp} =  \gamma \hspace{1in} h_{\parallel} = \gamma + L (h - \gamma) \ .
\end{equation}

\paragraph{Stability condition near symmetric fixed point.} This Hessian is circulant, so we can reuse a linear model result (see Appendix \ref{app:linear}) to immediately learn that
\begin{equation}
\boxed{ \ \rho_c(L) = \frac{1}{2 L} \left[ \frac{h_{\parallel}^2}{h_{\parallel} + \gamma} + (L - 1) \frac{h_{\perp}^2}{h_{\perp} + \gamma} \right] = \frac{1}{2 L} \left[ \frac{[ \gamma + L (h - \gamma) ]^2}{ L (h - \gamma) + 2 \gamma} + \frac{(L - 1)}{2} \gamma  \right] \ } \ .
\end{equation}
In the case of no regularization ($\gamma = 0$), we have the simpler critical arrival rate
\begin{equation}
\boxed{ \ \rho_c(L) = \frac{h}{2} = \frac{v}{2} W_*^{2 - 2/L} \ }  
\end{equation}
which asymptotically approaches $\rho_c(\infty) := (v/2) W_*^2$ as $L \to \infty$.

\clearpage

\section{More details on pooling parameters}
\label{app:pool}

In this appendix, we derive results related to the parameter-pooling model discussed in Sec. \ref{sec:pool}. It is defined in almost the same way as our linear model (see Sec. \ref{sec:linear} and Appendix \ref{app:linear}):
\begin{equation}
\frac{d}{dt} \vec{w} = \vec{r} \odot \frac{1}{ \rho} [ \vec{S} (\vec{w}^* - \vec{w}) + \vec{\xi} ] - \eta \gamma \vec{w} + \eta k \vec{D} \vec{w} \ .
\end{equation}
The main difference is that there is an additional term $\eta k \vec{D} \vec{w}$, where $\vec{D}$ is a `diffusion' matrix that implements a sort of spatial parameter pooling. In particular, $\vec{D}$ is defined such that
\begin{equation}
\sum_j D_{i j} w_j := w_{i+1} - 2 w_i + w_{i - 1}
\end{equation}
for all $i$. The interpretation is that information flows from nodes $i+1$ and $i-1$ to node $i$, and from node $i$ to nodes $i+1$ and $i-1$. 

For simplicity, we assume periodic boundary conditions, so that $\vec{D}$ is circulant rather than tridiagonal. Under this assumption, its eigenvalues are
\begin{equation}
d_q := - 4 \sin^2\left( \frac{\pi q}{N}  \right) \leq 0
\end{equation}
where $q = 0, 1, ..., N-1$, and it can be diagonalized using discrete Fourier basis functions. 

The derivation of mean and covariance dynamics for this model is almost the same as it was for the linear model studied in Appendix \ref{app:linear}; everything is the same up to the identification 
\begin{equation}
\gamma \vec{I} \to \gamma \vec{I} - k \vec{D} \ .
\end{equation}
For this reason, we omit the derivation and directly state the results. As in Appendix \ref{app:linear}, we assume that $\vec{S}$ is circulant for analytic tractability. This assumption is particularly important here because it implies that $\vec{S}$ and $\vec{D}$ commute.

\paragraph{Mean dynamics.} Mean dynamics follow
\begin{equation}
\boxed{ \ \frac{d}{dt} \bar{\vec{w}} = \eta \left[ \vec{S} (\vec{w}^* - \bar{\vec{w}})  - (\gamma \vec{I} - k \vec{D}) \bar{\vec{w}} \right]  \ } \ .
\end{equation}
The corresponding fixed point is
\begin{equation}
\bar{\vec{w}}^{\text{FP}} := ( \vec{S} + \gamma \vec{I} - k \vec{D})^{-1} \vec{S} \vec{w}^* \ .
\end{equation}
In the discrete Fourier basis, this becomes
\begin{equation}
w_q^{\text{FP}} := \frac{s_q}{s_q + \gamma - k d_q} w^*_q = B_q w^*_q \ .
\end{equation}
where $s_q$ denotes the $q$-th eigenvalue of $\vec{S}$ and $d_q$ denotes the $q$-th eigenvalue of $\vec{D}$. As noted in the main text, the specific form of $d_q$ makes the overall associated prefactor $B_q$ fairly spatial-frequency-dependent.

\paragraph{Covariance dynamics.} Covariance dynamics follow
\begin{equation}
\boxed{ \ \frac{1}{\eta} \frac{d}{dt} \vec{C} = - 2 \gamma \vec{C} -  \vec{C} \vec{S} + k \vec{C} \vec{D} + k \vec{D} \vec{C} -  \vec{S} \vec{C} + \frac{1}{\rho}  \sigma^2_{\text{obs}} \vec{I} + \frac{1}{\rho} \ \text{diag}( \bar{g}_i^2 + [ \vec{S} \vec{C} \vec{S} ]_{i i}  ) \ } \ .
\end{equation}
Moreover, the diagonal elements of the discrete-Fourier-transformed matrix $\tilde{\vec{C}} := \vec{F}^{\dag} \vec{C} \vec{F}$ (see Appendix \ref{app:linear} for more context) satisfy
\begin{equation} 
\frac{1}{\eta} \frac{d}{dt} \tilde{C}_{qq} = - 2 \gamma \tilde{C}_{qq} + 2 k d_q \tilde{C}_{q q} - 2 s_q  \tilde{C}_{q q} + \frac{1}{\rho}  \sigma^2_{\text{obs}}  + \frac{1}{\rho} \ ( A_{q q} +\frac{1}{N} \sum_{q'} s_{q'}^2 \tilde{C}_{q' q'}  )  \ .
\end{equation}
Following a slight modification of the argument from Appendix \ref{app:linear}, we obtain the stability condition
\begin{equation}
\boxed{ \ \rho > \frac{1}{2  N} \sum_q \frac{s_q^2}{s_q + \gamma - k d_q }  \ } \ .
\end{equation}

\clearpage

\section{More details on averaging over gradient arrivals}
\label{app:avggrad}

In this appendix, we analyze the mean and covariance dynamics of the gradient-accumulation model discussed in Sec. \ref{sec:avggrad} of the main text. It is defined similarly to the linear model (see Sec. \ref{sec:linear} and Appendix \ref{app:linear}), but with a `gradient accumulation' variable $z_i$ for each weight $w_i$:
\begin{equation}
\begin{split}
\frac{d}{dt} \vec{z} &= \vec{r} \odot \frac{1}{\eta \tau_z \rho} [ \vec{S} (\vec{w}^* - \vec{w}) + \vec{\xi} ]  - \frac{1}{\tau_z} \vec{z} \\
\frac{d}{dt} \vec{w} &= \eta ( \vec{z} - \gamma \vec{w} ) \ .
\end{split}
\end{equation}
This model's master equation has the form
\begin{equation}
\begin{split}
\frac{\partial P}{\partial t} =& \   - \eta \frac{\partial}{\partial \vec{w}} \cdot \left[  ( \vec{z} - \gamma \vec{w} ) P  \right] + \frac{1}{\tau_z} \frac{\partial}{\partial \vec{z}} \cdot ( \vec{z} P )\\
& + \ \eta \rho \sum_i  \int_{-\infty}^{\infty} d\xi  \ \mathcal{N}(\xi; 0, \sigma^2_{\text{obs}}) \ 
\left[ P(\vec{z} - \frac{1}{\eta \rho \tau_z} [g_i(\vec w)+\xi]\vec{e}_i, \vec{w} | t) - P(\vec{z}, \vec{w} | t) \right] \ .
\end{split}
\end{equation}
In this section, because there are multiple covariance matrices, we will use different notation than elsewhere. Let $\vec{Z} := \text{Cov}(\vec{z})$, $\vec{X} := \text{Cov}(\vec{z}, \vec{w})$ denote the cross-covariance, and $\vec{W} := \text{Cov}(\vec{w})$. 

The calculations required to treat this model largely resemble those from Appendix \ref{app:linear}, but are lengthier, and hence more tedious. Because they are straightforward, we skip a number of intermediate steps and mostly present the final results.

\paragraph{Mean dynamics.} This model's average dynamics are
\begin{equation}
\begin{split}
\frac{d}{dt} \vec{z} &= \frac{1}{\tau_z}  \vec{S} (\vec{w}^* - \bar{\vec{w}})   - \frac{1}{\tau_z} \bar{\vec{z}} \\
\frac{d}{dt} \bar{\vec{w}} &= \eta ( \bar{\vec{z}} - \gamma \bar{\vec{w}} ) \ .
\end{split}
\end{equation}
In particular, the steady state value of $\bar{\vec{w}}$ is the same as it is in the linear model; this kind of gradient accumulation does not produce bias, at least asymptotically.

\paragraph{Covariance dynamics.} After a tedious calculation, one can find that
\begin{equation}
\begin{split}
\frac{d}{dt} \vec{Z} &= - \frac{2}{\tau_z} \vec{Z} - \frac{1}{\tau_z} ( \vec{X} \vec{S} + \vec{S} \vec{X}^T )   + \frac{1}{\eta \rho \tau_z^2} \text{diag}[ g_i(\bar{\vec{w}})^2 + [\vec{S} \vec{W} \vec{S}]_{ii} + \sigma^2_{\text{obs}}  ] \\
\frac{d}{dt} \vec{X} &= \eta \vec{Z}- \left( \eta \gamma + \frac{1}{\tau_z} \right) \vec{X} - \frac{1}{\tau_z} \vec{S} \vec{W} \\
\frac{d}{dt} \vec{W} &=  \eta \left[ \vec{X} + \vec{X}^T - 2 \gamma \vec{W}  \right] \ .
\end{split}
\end{equation}

\paragraph{Diagonal $S$.} Assume $\vec{S}$ is diagonal and let $Z_i$, $X_i$, and $W_i$ denote the diagonals of these covariance matrices. The diagonal entries have dynamics
\begin{equation}
\begin{split}
\frac{d}{dt} Z_i &= - \frac{2}{\tau_z} Z_i - \frac{2}{\tau_z} s_i X_i   + \frac{1}{\eta \rho \tau_z^2} [ g_i(\bar{\vec{w}})^2 + s_i^2 W_i + \sigma^2_{\text{obs}}  ] \\
\frac{d}{dt} X_i &= \eta Z_i - \left( \eta \gamma + \frac{1}{\tau_z} \right) X_i  - \frac{1}{\tau_z} s_i W_i \\
\frac{d}{dt} W_i &=  2 \eta \left[ X_i -  \gamma W_i \right] \ .
\end{split}
\end{equation}
The stability of this three-dimensional system of ODEs is characterized by the $3 \times 3$ matrix
\begin{equation}
\vec{M} := \begin{pmatrix}
- \frac{2}{\tau_z} & - \frac{2 s_i}{\tau_z}  & \frac{s_i^2}{\eta \rho \tau_z^2}\\
\eta &  -  \left( \eta \gamma + \frac{1}{\tau_z} \right)  & - \frac{s_i}{\tau_z} \\
0 & 2 \eta  &  - 2 \eta \gamma
\end{pmatrix} \ .
\end{equation}
We can directly compute that its determinant is
\begin{equation}
\begin{split}
& \ - \frac{2}{\tau_z} \left[ 2 \eta \gamma ( \eta \gamma + 1/\tau_z) + 2 \eta s_i/\tau_z  \right] + \frac{2 s_i}{\tau_z} \left[  - 2 \eta^2 \gamma \right] + \frac{s_i^2}{\eta \rho \tau_z^2} \left[ 2 \eta^2  \right] \\
=& \ - \frac{2}{\tau_z} \left[ 2 \eta \gamma ( \eta \gamma + 1/\tau_z) + 2 \eta s_i (1/\tau_z + \eta \gamma   )  \right] + \frac{s_i^2}{\eta \rho \tau_z^2} \left[ 2 \eta^2  \right] \\
=& \ - \frac{4 \eta}{\tau_z} ( \eta \gamma + 1/\tau_z) \left[  \gamma  +  s_i  \right] + \frac{s_i^2}{\eta \rho \tau_z^2} \left[ 2 \eta^2  \right] \ .
\end{split}
\end{equation}
Stability requires that the determinant is negative. Hence, we need it to be true that
\begin{equation}
2( 1 + \eta \gamma \tau_z ) (  \gamma  +  s_i  ) - \frac{s_i^2 }{\rho}  > 0 \ .
\end{equation}
Equivalently,
\begin{equation}
\boxed{ \ \rho > \rho_c^i := \frac{s_i^2 }{2( 1 + \gamma \eta \tau_z) ( s_i + \gamma )} \ } \ .
\end{equation}
In the infinitely fast integration limit ($\tau_z \to 0$), this reduces to the linear model result (see Appendix \ref{app:linear}).

The corresponding steady state weight variances are
\begin{equation}
\boxed{ \ W_{i i} = \frac{1}{\rho - \rho_c^i} \frac{\sigma^2_{\text{obs}} + \bar{g}_i^2}{2 (s_i + \gamma) ( 1 + \gamma \eta \tau_z)} \ } \ .
\end{equation}
Interestingly, this expression is the same as what we found for the linear model, except that $\rho_c^i$ has a different form.

\paragraph{Circulant $S$.} If $\vec{S}$ is instead circulant, we must go through a lengthier version of the circulant algebra from Appendix \ref{app:linear}. In the end, we find almost the same results. Most importantly, we find a stability condition
\begin{equation}
\boxed{ \ \rho > \rho_c := \frac{1}{2 N (1 + \eta \gamma \tau_z)} \sum_k \frac{s_k^2}{s_k + \gamma}    \ }
\end{equation}
which reduces to the corresponding linear model condition in the $\tau_z \to 0$ limit.


\end{document}